\documentclass[aps,prc,preprint]{revtex4}
\usepackage{psfig}
\usepackage{epsfig}
\usepackage{graphics}
\usepackage{amsmath}
\tighten

\renewcommand{\vec}[1]{\mathbf{#1}}

\begin{document}
\bibliographystyle{revtex}

\title
{Deformation effects in low-momentum distributions
of heavy nuclei}

\author{
V.O. Nesterenko$^{1,2}$, V.P. Likhachev$^2$, P.-G. Reinhard$^{3}$, \\
J. Mesa$^2$,
W. Kleinig$^{1,4}$, J.D.T. Arruda-Neto$^{2,5}$, and A. Deppman$^2$}

\address
{$^{1}$ Bogoliubov Laboratory of Theoretical Physics,
Joint Institute for Nuclear Research
141980, Dubna, Moscow Region, Russia,
E-mail: nester@thsun1.jinr.ru}

\address{
$^2$ Instituto de F\'{i}sica, Universidade de
S\~{a}o Paulo, S\~{a}o Paulo, Brazil}

\address{
$^4$ Institut f\"ur Theoretische Physik,
Universit\"at Erlangen, D-91058, Erlangen, Germany}
\address{
$^4$ Technische Universit\"at Dresden,
Institut f\"ur Analysis, Dresden, D-01062, Germany}

\address{
$^5$
Universidade de Santo Amaro, S\~{a}o Paulo, Brazil}

\begin{abstract}
Momentum distributions (MD) of deep hole proton states in
$^{238}$U are studied paying particular attention to the
influence of deformation.  Two essentially different
mean-field models, Woods-Saxon (WS) and Skyrme-Hartree-Fock
with the SkM$^*$ force, are used. Noticeable deviations between
the WS and SkM$^*$ results are found. They are mainly due to
the difference in effective nucleon mass.
In particular, SkM$^*$ gives much weaker
deformation effects at low momenta than WS.
It is shown that, in spite of the deformation mixing,
MD at low momenta can serve for identification
of $K^{\pi}=1/2^+$ hole states originating from $s_{1/2}$
spherical sub-shells. Moreover, following the WS calculations,
the deformation results in additional $K^{\pi}=1/2^+$ states with
strong $l=0$ contributions. A possibility to probe such states
in knock-out experiments is discussed.
\end{abstract}

\pacs{21.60.-n, 25.85.Ge, 27.90.+b}

\maketitle


The momentum distribution (MD) of nucleons in a nucleus is an
important issue for analyzing single-particle aspects of nuclear
structure, for a review see \cite{Ant}.  One of the still open
problems is the influence of global nuclear deformation on MD.
There exists already some studies on that subject,
see, e.g. \cite{guer} and references. therein. Nonetheless,
the present knowledge about deformation effects in MD is
still rather poor, especially in heavy deformed nuclei.
At the same time, these effects can be essential
for a correct treatment of knock-out reactions, like $(e,e'p)$,
in rare earth and actinide regions and for investigation of
deep hole states in deformed nuclei (see, e.g. \cite{Likh}).

Experimental data on deep hole states are scarce in spherical
nuclei and practically absent in heavy deformed nuclei \cite{Ant}.
The deformation mixes spherical configurations in single-particle
wave functions \cite{Gar} and thus should influence the MD. At
the first glance, this effect, together with a dense spectrum in
deformed nuclei and possible coupling
with vibrations, should make MD too vague to extract from them
a reliable information about single-particle levels.
However, as is shown below, there remains a chance to identify
some hole $K^{\pi}=1/2^+$ levels through their MD at low momenta.
Moreover, the deformation mixing can increase
the number of such levels. Experimental observation of the
$K^{\pi}=1/2^+$ level sequence in  $(e,e'p)$ could be a
very important step in our knowledge on the mean field in
heavy deformed nuclei. Besides, it could have a serious impact
in such urgent problems, as, e.g., a search for shell-stabilized
super-heavy nuclei \cite{Afa}.

The present paper pursues two aims. First, we will try to exhibit,
at least briefly, a general view of deformation effects in deep hole
MD  of heavy deformed nuclei. To make the study more reliable,
two different single-particle potentials, phenomenological Woods Saxon
and  self-consistent Skyrme, will be used.
Second, we will demonstrate that low momentum distributions in heavy
deformed nuclei represent a relevant test for different models and
deserve a careful experimental investigation.


The calculations have been performed for $^{238}$U, a typical heavy
deformed nucleus. Two different models for single-particle structure
were used: the phenomenological Woods-Saxon (WS) potential
\cite{Pash,Ivan} with parameters from \cite{WS99} and the
self-consistent Skyrme-Hartree-Fock potential \cite{Skyrme} with the
parametrization SkM$^*$ \cite{skms}.  We have chosen SkM$^*$
as a widely established standard well suited to describe
deformed nuclei (SkM$^*$ was fitted to describe fission barriers in
actinide region). The WS calculations were performed in a basis
expansion while the SkM$^*$ calculations employed an axially symmetric
coordinate-space grid.  For both WS and SkM$^*$, the equilibrium
ground-state deformations were determined by minimizing the energy
of the system. MD were defined as single-particle densities
$n_\alpha({\vec k})=|\tilde\varphi({\vec k})|^2$
in momentum space. They were obtained by Fourier transforming the
spatial wavefunctions, for procedure and definitions see
\cite{Likh}. It is worth noting that in general single-particle models
are not well suited to describe MD because of the important
contributions from short- and long-range correlations \cite{Ant}.
However
these perturbing effects take place mainly in the high-momentum
domain with
$k > k_{\rm F}\approx 1.3 \; {\rm fm}^{-1} \approx 260\,{\rm MeV/c}$.
We will focus on MD in the low $k$
domain, which can be still described by single-particle models.


%
\begin{table}
\caption{\label{tab:tab1}
Quadrupole moments ($Q_2$), Fermi energy ($E_F$)
and the lowest ($E_0$) single-particle levels in $^{238}U$.
Experimental estimations for the quadrupole moment in
$^{238}$U lie in the interval $Q_2=11.1 - 11.3 $b
\protect\cite{NDS,Sol}.
}
\begin{center}
\begin{tabular}{|c|c|c|c|}
\hline
 Potential & $Q_2$[b] & $E_F$[MeV] & $E_0$[MeV] \\
\hline
 WS    & 11.0 & -7.5 & -33.7 \\
 SkM$^*$  & 11.1 & -6.2 & -39.8  \\
\hline
\end{tabular}
\end{center}
\end{table}
As is seen from table \ref{tab:tab1}, both single-particle models
reproduce nicely the ground state deformation (quadrupole moments) in
$^{238}$U.  The Fermi
energy $E_F$ is also nearly the same in both models.  However, the
lowest proton hole state $E_0$ is much deeper bound for SkM*.
The difference in $E_0-E_F$ indicates a different level density.
This is also visible in the single proton levels shown in table
\ref{tab:tab2}. The SkM$^*$ spectrum is less dense because of the
lower effective mass $m^*/m=0.79$ as compared to $m^*/m=1$ in the WS
model. Note that table \ref{tab:tab2} shows only a
subset of the levels whose MD will be discussed further in more detail.
\begin{table}
\caption{\label{tab:tab2}
Energies of selected proton hole states in $^{238}\mathrm{U}$,
calculated with WS and SkM$^*$ potentials. The lower part
of the table collects all the hole $K^{\pi}_i=1/2^+$ states.
For every state the spectroscopic quantum numbers of the spherical
ancestor are given. The low index in $K^{\pi}_i$
numerates states with a given $K^{\pi}$ from the bottom of the wall.
}
\begin{center}
\begin{tabular}{|c|c|c|c|}
\hline
$K^{\pi}_i[Nn_z\Lambda ]$ & nlj & \multicolumn{2}{|c|}{E, MeV}\\
\cline{3-4}
& & \; WS \;& \;SkM$^*$\; \\
\hline
$1/2^-_8[541]$ & $1h_{11/2}$ & -10.4 &  -9.5
    \\
$3/2^+_4[422]$ & $1g_{7/2}$ & -14.3 & -14.7
    \\
$1/2^-_6[301]$ & $2p_{1/2}$ & -15.5 & -16.4
    \\
$5/2^-_1[312]$ & $1f_{7/2}$ & -21.3 & -23.3
    \\
$1/2^-_4[321]$ & $1f_{5/2}$ & -21.6 & -23.9
    \\
\hline \hline
  $1/2^+_{10}[660]$ &$1i_{13/2}$ & -8.3 & -6.8 \\
  $1/2^+_9[400]$ &$3s_{1/2}$ & -7.2 & -6.9 \\
  $1/2^+_8[411]$ & $2d_{3/2}$ &  -9.8 & -9.7 \\
  $1/2^+_7[420]$ &$2d_{5/2}$& -13.1 & -13.5 \\
  $1/2^+_6[431]$ &$1g_{7/2}$ & -16.4 & -17.1 \\
 $1/2^+_5[440]$ & $1g_{9/2}$& -19.3 & -20.5 \\
 $1/2^+_4[200]$ & $2s_{1/2}$ & -22.8 & -25.4 \\
 $1/2^+_3[211]$ & $1d_{3/2}$ & -26.0 & -29.5 \\
 $1/2^+_2[220]$ &$1d_{5/2}$ & -28.1 & -32.1 \\
 $1/2^+_1[000]$ & $1s_{1/2}$ & -33.7 & -39.8 \\
\hline
\end{tabular}
\end{center}
\end{table}

Figure 1
shows the calculated MD for six deep hole
proton states given in the table \ref{tab:tab2}. These states
represent a selection with different energies and quantum numbers.
SkM$^*$ results are given for the equilibrium deformed ground state
and, for reasons of comparison, in the spherical limit produced by
constraint. One sees that in both these cases MD are very close.
Modest changes are caused by the deformation at low momenta
in $1/2^-[321]$, $1/2^+[440]$, and $1/2^-[541]$,
i.e. in states with a minimal projection $K$.  At the same time, as
illustrated by the plots for $5/2^-[312]$ and $3/2^+[422]$,
MD of the hole states with $K\ge 3/2$ are almost unaffected.
The latter can be easily explained by the fact that states with high
$K$ have lower level density and so less chances for deformation
mixing.

Figure 1
compares also SkM$^*$ and WS results.
Unlike SkM$^*$, WS gives impressive deformation effects at
low momenta for the $1/2^-[321]$, $1/2^+[440]$, $3/2^+[422]$,
and $1/2^-[541]$ states, where one-bump spherical MD are transformed
by deformation to more complex structures.
The discrepancies between WS and SkM$^*$
can be explained, to a large extent, by the lower effective mass
$m^*/m=0.79$ in SkM$^*$. This stretches the single-particle spectrum
\cite{Mahaux} for SkM$^*$ and thus weakens the deformation effects.

As was mentioned above, single-particle models are not reliable in the
high-momentum domain with $k > 260\,{\rm MeV/c}$.  Nevertheless, it is
worth to comment on some basic trends of the models as such, hinted by
high-momentum MD. First, as compared with SkM$^*$, the MD of WS are
characterized by a stronger high-momentum tail and thus by a smaller
peak height at $k\sim 170$ MeV/c. Second, SkM$^*$ results for
$1/2^+[440]$ and $1/2^-[541]$ show that the deformation can influence
the high-energy MD tail.

Figure 1
shows that MD  for  $1/2^+[440]$ is
peaked at $k=0$ thus manifesting an appreciable $l=0$
component. This feature can essentially facilitate
identification of the hole $K^{\pi}=1/2^+$ levels in
knock-out reactions. We are thus inspecting these states
in more detail. Figure 1
collects MD
for all hole $K^{\pi}=1/2^+$ levels in $^{238}$U (the
spectroscopic information is given in table \ref{tab:tab2}).
It is seen that WS calculations predict an appreciable $l=0$
strength (peak at $k=0$) for 7 states, namely for all those
with $\Lambda =0$. The maximal strength takes place for states
$1/2^+[000]$, $1/2^+[200]$, and $1/2^+[400]$ originating from
$1s$, $2s$, and $3s$ spherical subshells, respectively. The other four
states, $1/2^+[220]$, $1/2^+[440]$, $1/2^+[420]$, and $1/2^+[660]$,
have higher $l$ ancestors but acquire such a feature due to the
admixture of $l=0$ spherical sub-shells, caused by the deformation
mixing. So, the deformation leads to an increased number
of $K^{\pi}=1/2^+$ states with a strong $l=0$ component. As was mentioned
above, SkM$^*$ produces much weaker deformation effects.  Indeed,
SkM$^*$ gives $k=0$  peaks only for the states originating from $1s$,
$2s$, and $3s$ spherical subshells, being consistent in this case with
the WS results. The $l=0$ strength for other states is negligible.

Altogether, figure 2
indicates that MD at low
momenta allow to extract important information on the
underlying mean field. Namely, a sequence
of hole $K^{\pi}=1/2^+$ levels can be found experimentally.
The question is whether one can resolve them.
In the following, we will analyze this point in more detail.

Figure 3
shows the calculated energy distribution
of MD strength at $k=0$  for the hole $K^{\pi}=1/2^+$ levels.
It is seen that SkM$^*$ and WS  represent two different
scenarios.
In the SkM$^*$ case, only $1/2^+[000]$, $1/2^+[200]$, and $1/2^+[400]$
levels originating from $1s$, $2s$, and $3s$ spherical subshells and
giving the biggest $k=0$ peaks can be resolved. The picture looks like
in heavy spherical nuclei. Indeed, in the $^{208}{\rm Pb}(e,e'p)$ reaction,
the $l=0$ strength was observed only as the ground state $3s_{1/2}$ and deep
hole $2s_{1/2}$ structures \cite{Heyde}. The former was found as a strong
narrow peak and the latter as a broad bump spreaded in the interval
7-8 MeV with the centroid at $\sim 21.3$ MeV (the $1s_{1/2}$ was beyond
the energy interval covered in the experiment). WS calculations predict
more interesting situation when at least three
additional levels ($1/2^+[440]$, $1/2^+[420]$,
and $1/2^+[660]$) placed between $1/2^+[200]$ and $1/2^+[400]$
have a chance to be observed (more deeply bound levels can
hardly be indicated because of a too strong broadening expected
for them).

Single particle energies in mean field models have well
defined discrete values as indicated by the simple bars in
figure 3.
However, $l$-strengths
measured in $(e,e'p)$ experiments are spreaded,
becoming broader while going away from the Fermi
energy towards deeply bound states.
The spread is caused by correlations
beyond the mean field approximation, in particular by
the coupling the holes with 1ph configurations and vibrations
(see \cite{Ant,Heyde,Mahaux} and references therein).
The deformation also contributes to the spread.
Knock-out reactions are sensitive only to single-hole
components which are the door-way states for them. So, the
MD signal at $k=0$ will not be masked by complex components of
the wave function, caused by the correlations.
But the signal will be smeared out energetically to
a more or less broad bump corresponding to a certain $K^{\pi}=1/2^+$
level.  The question is whether these bumps overlap each other or they
are still  separated enough to allow identification of the various
$K^{\pi}=1/2^+$ levels in  $(e,e'p)$ reaction.

As is seen from figure 3,
in the SkM$^*$ calculations
the relevant levels $1/2^+[200]$ and $1/2^+[400]$  are well
separated in energy and so, in spite of a (weak) deformation
mixing, can be observed. In the WS calculations, the levels
$1/2^+[200]$, $1/2^+[440]$, $1/2^+[420]$, $1/2^+[660]$,
and $1/2^+[400]$ are separated by the intervals 3-6 MeV
(which hint at the energy scale for the deformation mixing).
Following the $^{208}{\rm Pb}(e,e'p)$ data \cite{Heyde},
the correlation spread should not exceed 7-8 MeV and for
most of the states (as less bound) should even much weaker.
For these states the correlation and deformation spreads
seem to be of the same order of magnitude. So, we probably
have an interesting situation when the deformation does not
noticeably hinder the identification of $l=0$ structures
but instead can increase the number of such structures.


In summary, momentum distributions (MD) for single-hole proton states
in $^{238}$U were analyzed within two different mean field models:
phenomenological Woods-Saxon (WS) and self-consistent
Skyrme-Hartree-Fock with the force
SkM$^*$. WS gives much stronger deformation effects
than SkM$^*$, which is explained by the lower effective
mass in SkM$^*$ (in fact SkM$^*$ description stays
close to the spherical case). The deformation mainly influences
$K^{\pi}=1/2^+, 1/2^-$ hole states in the low momentum domain.
WS results predict a sequence of $K^{\pi}=1/2^+$ states
with strong $l=0$ contributions.
Some of these states acquire the
$l=0$ strength due to the deformation mixing.
The property of $l=0$ strength to give a peak in MD at
zero momenta considerably simplifies identification
of $l=0$ structures. Our analysis shows
that the $K^{\pi}=1/2^+$ states have a good chance to
be discriminated in knock-out reactions. Observation of a sequence
of such states in heavy deformed nuclei
would provide a valuable information about their mean field.
Besides, experimental study of the low-momentum MD
in heavy deformed nuclei would help to discriminate the various
mean-field models.

\bigskip

\noindent
Acknowledgment:
The work was partly supported by grants from
 FAPESP (Grant No. 2001/06082-1), RFBR (Grant No. 00-02-17194),
and Germany-BLTP JINR ( Heisenberg-Landau).
We express also our gratitude to Profs. A.N. Antonov,
A.V. Afanasjev and J.R. Marinelli for their useful
comments.


\newpage

{\bf \large FIGURE CAPTIONS}

\vspace{0.5cm}\indent
{\bf Figure 1}:
MD for selected hole proton states in $^{238}$U:
SkM$^*$ (solid curve) and WS (dashed curve) results
at equilibrium deformations as well as SkM$^*$ results
in the spherical limit (dotted curve).

\vspace{0.5cm}\indent
{\bf Figure 2}:
MD for $K^{\pi}_i=1/2^+$ hole proton states in $^{238}$U:
SkM$^*$ (solid curve) and WS (dashed curve) results.

\vspace{0.5cm}\indent
{\bf Figure 3}:
MD at $k=0$ for the $K^{\pi}_i=1/2^+$ hole proton states
(the same as in figure 2).

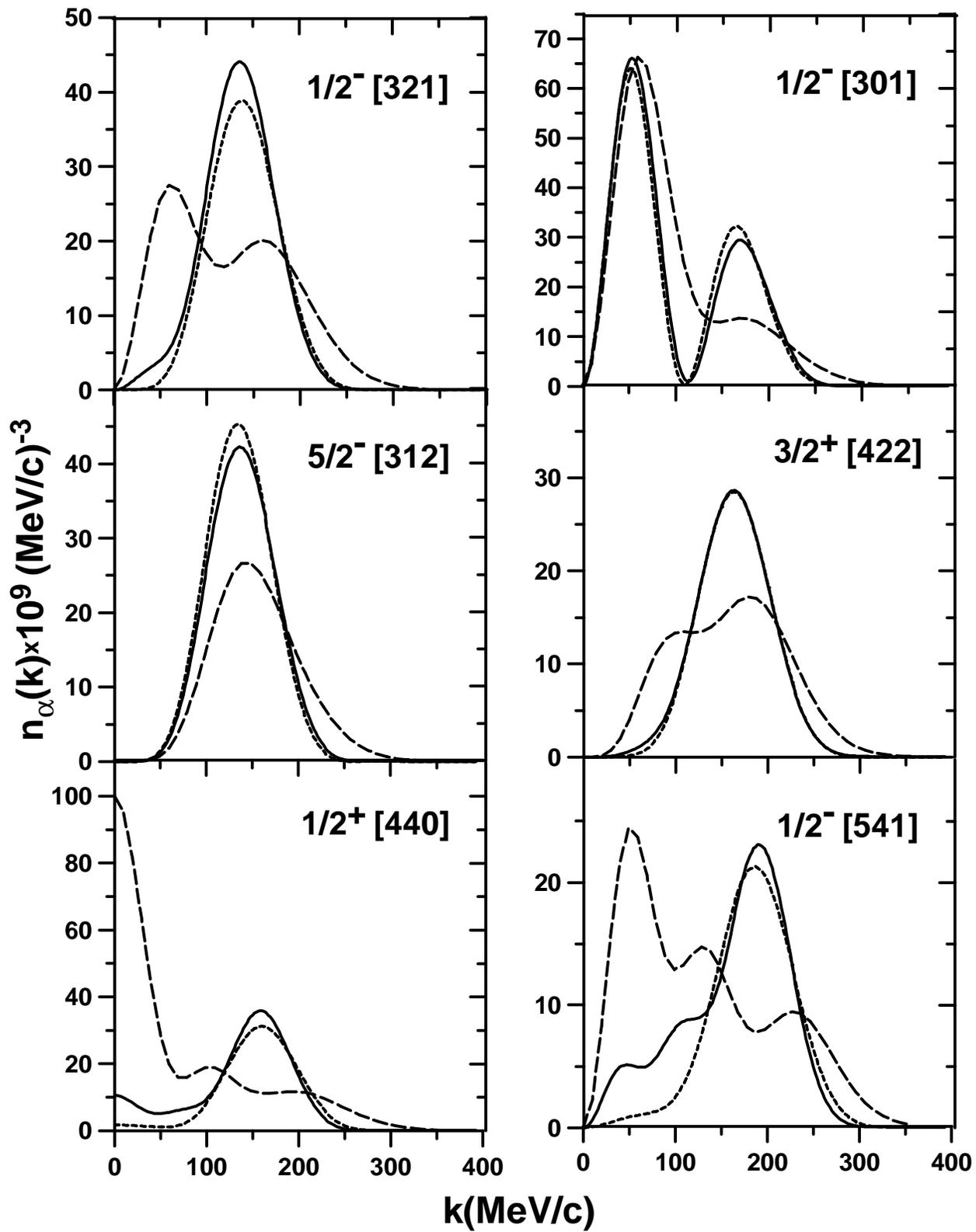
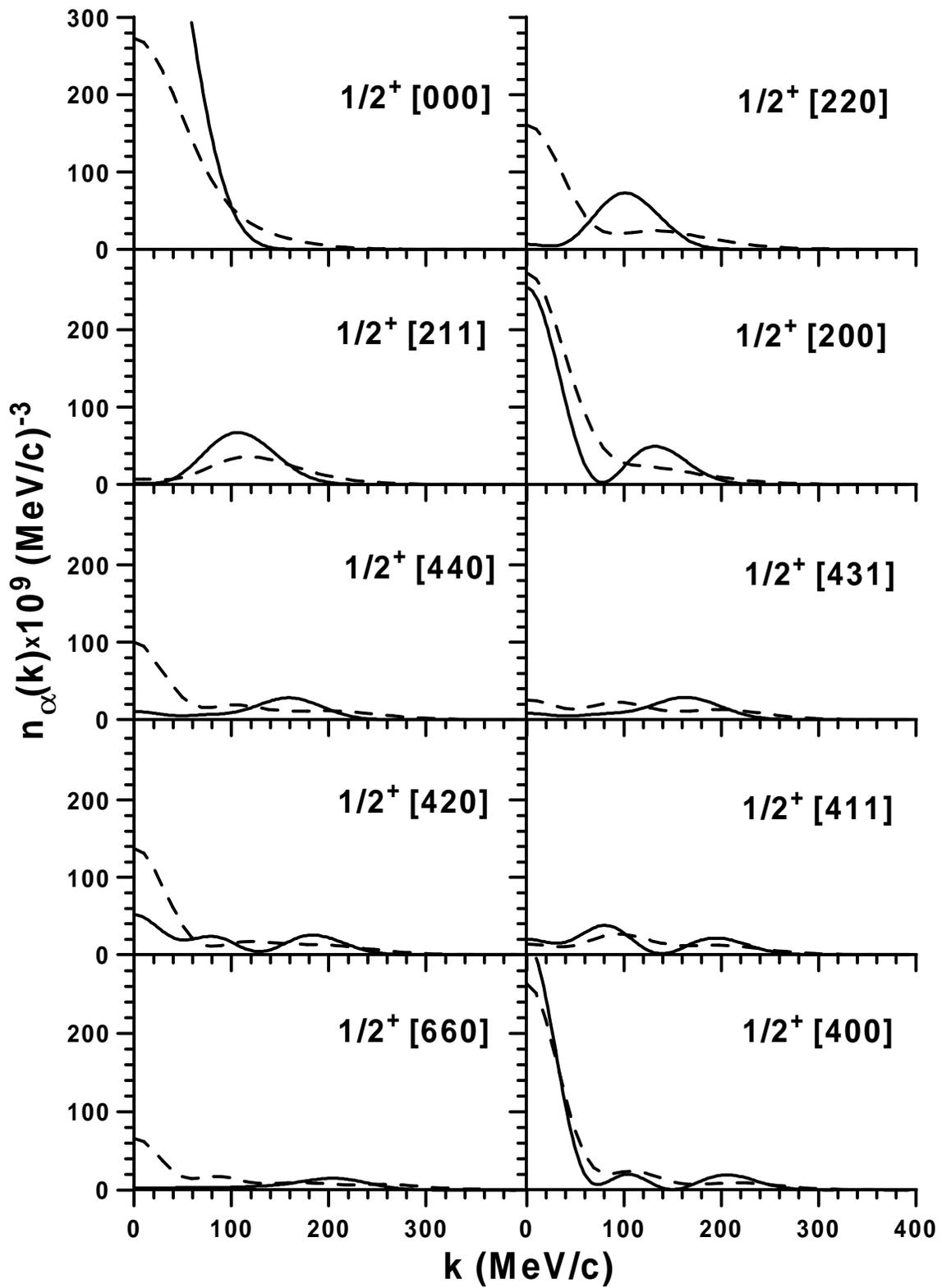
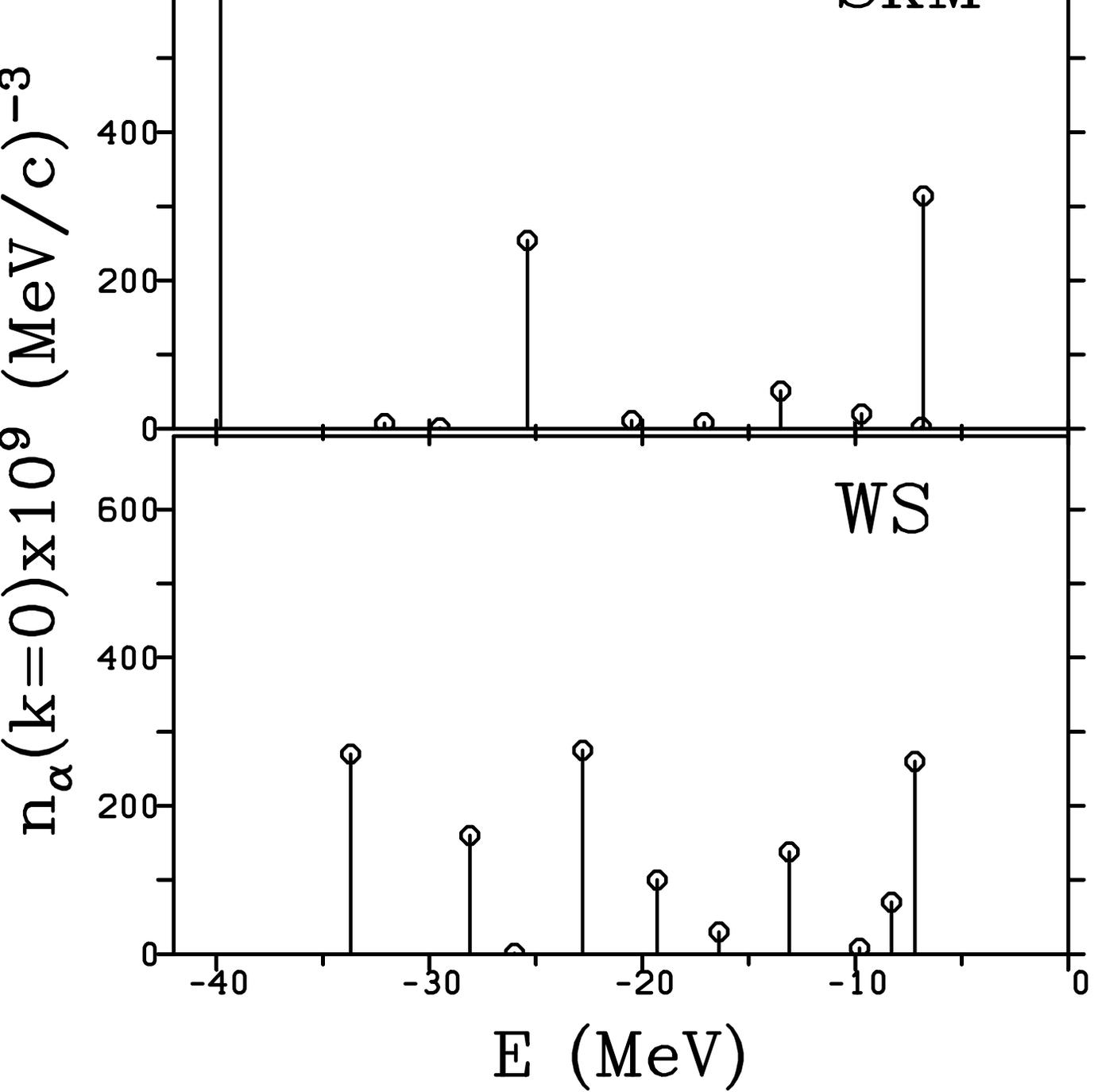
\end{document}